\documentclass[apj,numberedappendix]{emulateapj}
\usepackage{amsmath}
\usepackage{color}
\usepackage{amsmath}
\shorttitle{Polarization radiation from X-ray binaries}
\shortauthors{Zhang et al.}

\begin{document}
\title{Polarization radiation with turbulent magnetic fields from X-ray binaries}

\author{Jian-Fu Zhang\altaffilmark{1}, Fu-Yuan Xiang\altaffilmark{1} and  Ju-Fu Lu\altaffilmark{2}}
\altaffiltext{1}{Department of Physics, Xiangtan University, Xiangtan 411105, China;}
 \email{jfzhang@xtu.edu.cn;fyxiang@xtu.edu.cn;lujf@xmu.edu.cn}  %
\altaffiltext{2}{Department of Astronomy and Institute of Theoretical Physics and Astrophysics, Xiamen University, Xiamen, Fujian 361005, China}

\begin{abstract}
We study the properties of polarized radiation in turbulent magnetic fields from X-ray binary jets. These turbulent magnetic fields are composed of large- and small-scale configurations, which result in the polarized jitter radiation when the characteristic length of turbulence is less than the non-relativistic Larmor radius. On the contrary, the polarized synchrotron emission occurs, corresponding to a large-scale turbulent environment. We calculate the spectral energy distributions and the degree of polarization for a general microquasar. Numerical results show that turbulent magnetic field configurations can indeed provide a high degree of polarization, which does not mean that a uniform, large-scale magnetic field structure exists. The model is applied to investigate the properties of polarized radiation of black hole X-ray binary Cygnus X--1. Under the constraint of multiband observations of this source, our studies demonstrate that the model can explain the high polarization degree at MeV tail and predict the highly polarized properties at high-energy $\gamma$-ray region, and that the dominant small-scale turbulent magnetic field plays an important role for explaining the highly polarized observation at hard X-ray/soft $\gamma$-ray bands. This model can be tested by polarization observations of upcoming polarimeters at high-energy $\gamma$-ray bands.
\end{abstract}


\section{Introduction}
\label{Intro}
A compact X-ray binary consists of a stellar companion and a compact object, such as black hole or neutron star. Among Galactic X-ray binaries detected, there are about twenty microquasars that present the extended relativistic radio jets. It is widely considered that during the low/hard spectral state, radio through infrared (IR) emissions are from synchrotron processes of relativistic electrons. However, the origin of both X-ray and $\gamma$-ray emissions still remains open; they may be from the relativistic jet, hot accretion flow, and/or disk-corona region (e.g., \citealt{Markoff01,Markoff05,Yuan05,Peer09,Zhang13}). As far as we know, it seems to be difficult to distinguish them only by fitting emission spectra. We in this paper study the properties of polarization radiation of microquasars. The polarized radiation is an intrinsic property of the electromagnetic radiation, which carries important information on astrophysical sources, such as the geometry and orientation of the magnetic field, radiation mechanism. Therefore, polarization study is one of the important methods to probe an X-ray binary, leading to a better understanding of radiative mechanism, of configuration of magnetic fields, and of matter composition as well as energetics of the jet.

The polarization properties of emissions in X-ray binaries have been strongly studied at the radio waveband (e.g., \citealt{Fender06}), where a relatively low level of polarization is associated with synchrotron process in the jet. Similar to the studies for multi-wavelength spectral energy distributions (SEDs), it is more interesting to study multi-wavelength polarization properties (e.g., \citealt{Russell14} for a recent work on the microquasar Cygnus X--1). Recently, \cite{Laurent11} observed strong polarization, 67 $\pm$ 30$\%$, of the high-energy radiation of Cygnus X--1, between 400 keV and 2 MeV (so-called MeV tail), using the International Gamma-Ray Astrophysics Laboratory Imager on board the Integral satellite (INTEGRAL/IBIS). This result, i.e., the highly polarized emission at MeV tail, later was independently confirmed by using the INTEGRAL/SPI instrument, namely, 76 $\pm$ 15 $\%$ at 0.23--0.85 MeV (\citealt{Jourd12}). By separating the observations of Cygnus X--1 into hard, soft, and intermediate/transitional spectral states, \cite{Rodr15} confirmed that in the hard spectral state, where it is expected to be the presence of the jet, the degree of linear polarization is 75 $\pm$ 32$\%$ between 0.4 MeV and 2 MeV and the polarization angle is 40$^\circ$.0 $\pm$ 14$^\circ$.3.

Producing such a high polarization at MeV tail was claimed as the synchrotron emission of relativistic electrons in the jet due to a uniform (ordered), large-scale magnetic field configuration (\citealt{Laurent11,Jourd12,Russell14,Rodr15}). Alternatively, it has been proposed that a hot accretion flow, which requires mono-direction motion of relativistic electrons along highly ordered
magnetic fields in the inner regions of the accretion flow (\citealt{Veledina13}), or a hot, highly magnetized plasma corona around the black hole, where hadron and lepton interacting with matter and magnetic fields are invoked, may produce the highly polarized emission at MeV tail (\citealt{Romero14}).

In these works mentioned above, almost all investigations with regard to a high polarization degree usually are considered as the existence of a uniform, large-scale magnetic field configuration. In this context, one could intuitively understand the polarization properties by the following methods. In an optically thin region, the electric vector of the emitted radiation is perpendicular to the magnetic field direction and the linear polarization degree is given by $\Pi=\frac{p+1}{p+7/3}$ (\citealt{Rybicki79}), where $p$ is the spectral index of emitting relativistic electrons. However, the electric vector of the emitted radiation is parallel, rather than perpendicular, to the magnetic field direction in an optically thick region, where the degree of polarization is written as $\Pi=\frac{3}{6p+13}$ (\citealt{Longair11}). In this paper, one of our main purposes is to suggest that a highly polarized emission does not necessarily require that the magnetic field in X-ray binary jets have a uniform, large-scale configuration, if the configuration of turbulent magnetic fields
has a certain anisotropy.

As is known, magnetic turbulence is ubiquitous in astrophysical objects and plays an critical role in key astrophysical processes, such as star formation, acceleration and propagation of cosmic rays, heat transport, magnetic reconnection, amplification of magnetic fields, and accretion processes (see \citealt{Zhang16}, for a brief review in the introduction section). It is necessary for the existence of turbulent magnetic fields to induce a diffusive shock acceleration. When particles cross successively the shock front, magnetic turbulence would trap these particles and results in the particle energy gain. Moreover, particle-in-cell simulations demonstrate that a Weibel instability, which is a crucial ingredient for amplification of magnetic fields and production of collisionless shock waves (e.g., \citealt{Spit08,Medvedev11}), can generate turbulent magnetic fields in the shock waves, and that a conspicuous anisotropy of the magnetic turbulence appears at the saturation stage of field amplification (\citealt{Medvedev11}).

 We consider that random, turbulent magnetic fields appear in the jet of black hole X-ray binaries (see \citealt{Laing1980,Kelner13,Prosekin16} for theoretical basics). These magnetic fields mixed by large- and small-scale random structures are confined to a certain plane (slab), i.e., the limiting case of a three-dimensional geometry of the total compression, which results in the polarized synchrotron and jitter emissions. With this magnetic field configuration, we explore whether relativistic electrons in random fields can produce an expected highly polarized emission in the jets of X-ray binaries.

The paper is organized as follows. Model descriptions including relativistic electron evolution, model geometry and radiative mechanisms of polarization are presented in Section~2. Numerical results of the model are shown in Section~3. Section~4 is an application of the model to Cygnus X--1. Conclusions and discussion are presented in Section~5.

\section{Model Descriptions}
\label{model}

We employ a classical geometry of the microquasar in which dipole jets are launched from the inner regions of an accretion disk, perpendicular to the orbital plane of binary system. Thanks to collision interactions between the ejected matter, shock waves are generated in the bulk motion of the jets. Theoretical works suggested that the ordered, large-scale magnetic fields anchored in an accretion disk or a rotating black hole induce the generation of jets by means of a magneto-centrifugal mechanism \citep{BZ77,BP82}. Therefore, it is natural that when modelling the multi-broadband observations of X-ray binaries, the ordered, large-scale magnetic field configuration is usually assumed in operation in many published works. Due to the generation of the shock waves and compression between them, it is easy to envision that the large-scale magnetic fields could be disturbed or disrupted to bring about random, small- and large-scale fields, which are mixed by some random, small-scale turbulent magnetic fields generated behind shock fronts by the turbulence. The anisotropy of these turbulent fields can be generated by compression of shock waves within the jet or shear at the jet boundary layer of initially chaotic magnetic fields as well as by production of anisotropic turbulence by the shock wave itself. In this work, anisotropic structures are confined to a limiting case of the total compression, i.e., a slab geometry (e.g., \citealt{Laing1980}), in which directions of magnetic fields are parallel to the slab plane.

In these highly turbulent environments, when the characteristic scale of turbulence $\lambda$, which is of the order of the plasma skin-depth, is greater than the non-relativistic Larmor radius $R_{\rm L}=m_ec^2/eB$, radiation proceeds in the usual synchrotron regime. Here, $B$ is the strength of the magnetic field as a function of the height of the jet, and other parameters have a conventional meaning. On the contrary, $\lambda\ll R_{\rm L}$, the emission is referred to as diffusive synchrotron radiation (\citealt{Topt87}), or jitter radiation (\citealt{Medvedev00,Kelner13}). Below, we first study the evolution of relativistic electrons in jets, which has an isotropic distribution along the jets. Subsequently, the configurations of the magnetic slab and the jet and radiative mechanisms of polarization are presented.

\subsection{Relativistic Electron Evolution}
We consider that the evolution of steady-state relativistic electrons in a conical jet is formulated as \citep[][]{ZAA14,ZhangLu15,Zhang15}
\begin{equation}
\frac{1}{z^2}{\partial \over \partial z} [\Gamma_{\rm j} \beta_{\rm j}cz^2 \tilde{N}_{\rm \gamma}(\gamma,z)] + {\partial \over \partial
\gamma} \left[\Gamma_{\rm j} \beta_{\rm j}c\tilde{N}_{\rm \gamma}(\gamma,z) {d\gamma \over  dz}\right] = Q_{\rm in}(\gamma,z), \label{dNdz1}
\end{equation}
where the first term denotes spatial advection, corresponding to the divergence term, $\triangledown\cdot\mathbf{\upsilon} \tilde{N}_{\rm \gamma}$, in a spherical coordinate, and the second term energy losses of relativistic electrons. $Q_{\rm in}$ is called the source term, that is, the injection rate of relativistic electrons, which has the dimension ${\rm erg}^{-1}\ {\rm s}^{-1}\ {\rm cm}^{-3}$. $ \tilde{N}_{\rm \gamma}$ is the energy density of electrons, as a function of the electron energy $\gamma$ and  the jet height $z$ from the central compact object. $\Gamma_{\rm j}$ is the bulk Lorentz factor of the jet, and $\beta_{\rm j}=\sqrt{\Gamma_{\rm j}^2-1}/\Gamma_{\rm j}$ the bulk velocity. In order to make Equation (\ref{dNdz1}) more compact and numerical calculation convenient, we introduce a symbol $N_{\rm \gamma}$, then let $N_{\rm \gamma}(\gamma,z)=\tilde{N}_{\rm \gamma}\pi R_{\rm jet}^2\Gamma_{\rm j} \beta_{\rm j}c$ with dimension $\rm erg^{-}\ s^{-1}$. $R_{\rm jet}=z{\rm tan}\delta$ is the radius of the jet, where $\delta$ is a half-opening angle of the jet. In this way, Equation (\ref{dNdz1}) is rewritten as the following compact form \citep[see also,][]{Moderski03,ZhangLu15,Zhang15},
\begin{equation}
{\partial N_{\rm \gamma}(\gamma,z) \over \partial z}+{\partial \over \partial
\gamma} \left[N_{\rm \gamma}(\gamma,z) {d\gamma \over dz}\right] = \pi R_{\rm jet}^2 Q_{\rm in}(\gamma,z)
, \label{dNdz2}
\end{equation}
where, the energy loss rate of relativistic electrons along the jet is
\begin{equation}
{d\gamma \over dz} = {1 \over  c\beta_{\rm j} \Gamma_{\rm j}}\left(d\gamma \over dt{'}\right)_{\rm rad}- {2 \over
3}{\gamma \over z}. \label{dgdz}
\end{equation}
The total radiative loss rates of electrons, $(d\gamma/dt{'})_{\rm rad}$, include synchrotron and jitter emissions, and Comptonization of
the photons from the companion. In this work, we neglect synchrotron self-Compton scattering and Comptonization of
the photons from accretion disk and corona components, on the basis of the studies of \cite{Zhang14}. The second term of the right-hand side of Equation (\ref{dgdz}) indicates the adiabatic loss of relativistic electrons due to an adiabatically expanding jet.

The injection of the accelerated electrons is considered as
\begin{equation}
Q_{\rm in}(\gamma,z)=Q_{\rm 0}\gamma^{-p}{\rm exp}(-\gamma/\gamma_{\rm cut}),
\end{equation}
where $p$ and $\gamma_{\rm cut}$ are the spectral index and break energy of the relativistic electrons, respectively. The normalization constant of relativistic electrons, $Q_{\rm 0}$, is determined by
\begin{equation}
L_{\rm rel}=\pi\int^{z_{\rm max}}_{z_0}{\rm d}z\int^{\gamma_{\rm max}}_{\gamma_{\rm min}}{\rm d}\gamma R_{\rm jet}^2 Q_{\rm in}(\gamma,z) \gamma, \label{Lrel}
\end{equation}
where $L_{\rm rel}=\eta_{\rm rel}\eta_{\rm jet}L_{\rm acc}$ is the power of injection electrons, and $L_{\rm acc}=\dot{M}_{\rm acc} c^2$ is the accretion power of the system via stellar wind outflows of the companion star. Here, $\dot{M}_{\rm acc}\sim 10^{-8}M_{\odot}{\rm yr}^{-1}$ is the mass accretion rate for a high-mass system. The power of relativistic electrons, $L_{\rm rel}=\eta_{\rm rel}L_{\rm jet}$, is a fraction of the jet power of $L_{\rm jet}=\eta_{\rm jet}L_{\rm acc}$, where $\eta_{\rm jet}$ is set as 0.1 and $\eta_{\rm rel}$ is an adjustable parameter. $\gamma_{\rm min}$ and $\gamma_{\rm max}$($\sim10^7$) are minimum and maximum energies of the relativistic electrons, respectively.

\subsection{Model Geometry}
\label{MGeo}
\begin{figure}[!htb]
\begin{center}
\includegraphics[width=90mm,height=80mm,bb=212 149 503 408]{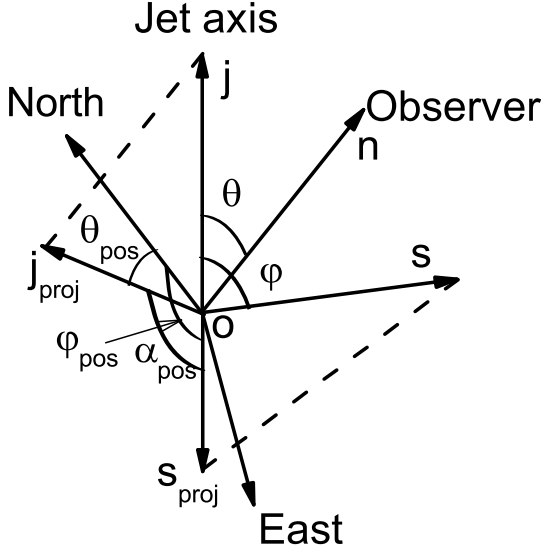}
\caption{A schematic illustration of the positions for the slab (with the unit normal vector \textbf{\emph{s}}) of turbulent magnetic fields compressed by shock waves and for the jet (a unit 3-D vector \textbf{\emph{j}}), as well as of the sky plane perpendicular to the line of sight, \textbf{\emph{n}}. The 2-D projections on the sky plane of the 3-D vectors \textbf{\emph{j}} and \textbf{\emph{s}} are denoted as \textbf{\emph{j}}$_{\rm proj}$ and \textbf{\emph{s}}$_{\rm proj}$, respectively. Note that three vectors \textbf{\emph{j}}, \textbf{\emph{n}}, and \textbf{\emph{s}} are not coplanar except for the case of a particular space position.}
\label{fig:diagram}
\end{center}
\end{figure}

In order to determine the positions of both the slab and the jet on the sky plane, and formulate the subsequent polarization radiative processes, we in Figure \ref{fig:diagram} provide a schematic illustration for the model geometry. The definition of the position angles of both the jet and slab is associated with the angles on the sky plane relative to the North celestial pole, and these angles are counted to the East, i.e., counter-clockwise. As shown, the direction of the line of sight is denoted as the unit vector \textbf{\emph{n}} that is perpendicular to the sky plane, the direction of the jet as the unit 3-D vector \textbf{\emph{j}}, and the normal direction to the slab of turbulent magnetic fields as the unit vector \textbf{\emph{s}}. In the three-dimensional coordinate reference frame constituted by both the sky plane and the observer line of sight, \textbf{\emph{s}} also is a unit 3-D vector. It should be noticed that generally three vectors \textbf{\emph{n}}, \textbf{\emph{j}}, and \textbf{\emph{s}} are not coplanar except for in the particular space position; the 2-D projection vectors on the sky plane of the two latters are indicated as \textbf{\emph{j}}$_{\rm proj}$, and \textbf{\emph{s}}$_{\rm proj}$, respectively. Now we distinguish the following completely different angles:

(1) The angle $\theta$ between the jet direction \textbf{\emph{j}} and the line of sight \textbf{\emph{n}}. Their scalar product is given by ${\rm cos}\theta=\textbf{\emph{j}}\cdot\textbf{\emph{n}}$.

(2) The angle $\theta_{\rm pos}$ between the projection of the jet $\textbf{\emph{j}}_{\rm proj}$ on the sky plane and the direction to the North celestial pole. Here, $\textbf{\emph{j}}_{\rm proj}=(\textbf{\emph{j}}-{\rm cos}\theta\textbf{\emph{n}})/|\textbf{\emph{j}}-{\rm cos}\theta\textbf{\emph{n}}|$ is the unit 2-D vector for the projection of the jet vector \textbf{\emph{j}} on the sky plane and $\hat{e}_N$ is the unit vector on the sky plane pointing to North. Further, we have ${\rm cos}(\theta_{\rm pos})=\textbf{\emph{j}}_{\rm proj}\cdot\hat{e}_N$, where $\theta_{\rm pos}$ is called the position angle of the jet.

(3) The angle $\varphi$ between the normal to the slab \textbf{\emph{s}} and the line of sight \textbf{\emph{n}}. Two unit vectors' scalar product is ${\rm cos}\varphi=\textbf{\emph{s}}\cdot\textbf{\emph{n}}=\sigma$. The symbol $\sigma$ is here introduced for the sake of compactness when writing radiative formulae.

(4) The angle $\varphi_{\rm pos}$ between projection of the normal to the slab $\textbf{\emph{s}}_{\rm proj}$ and the North. $\textbf{\emph{s}}_{\rm proj}=(\textbf{\emph{s}}-{\rm cos}\varphi\textbf{\emph{n}})/|\textbf{\emph{s}}-{\rm cos}\varphi\textbf{\emph{n}}|$ is the unit 2-D vector for the projection of the normal vector \textbf{\emph{s}} to the slab on the sky plane. Thus, we have ${\rm cos}(\varphi_{\rm pos})=\textbf{\emph{s}}_{\rm proj}\cdot\hat{e}_N$, where $\varphi_{\rm pos}$ is the position angle of the slab.

(5) The angle $\alpha_{\rm pos}$ between projection of the normal to the slab $\textbf{\emph{s}}_{\rm proj}$ and the projection of the jet $\textbf{\emph{j}}_{\rm proj}$. The scalar product is ${\rm cos}(\alpha_{\rm pos})=\textbf{\emph{j}}_{\rm proj}\cdot\textbf{\emph{s}}_{\rm proj}$.

(6) The polarization angle $\Phi$ between the polarization vector $\textbf{\emph{e}}$ and the North.

(7) The angle $\Psi$ between polarization (electric) vector of the radiation $\textbf{\emph{e}}$ on the sky plane and projection of the normal to the slab $\textbf{\emph{s}}_{\rm proj}$. $\Psi$ would appear in the radiation formulae of both synchrotron and jitter processes. In the case of linearly polarized radiation, if $\Psi=90^\circ$, radiative flux is $P_{\rm jit}=I_{\rm jit}-Q_{\rm jit}$ for jitter radiation and $P_{\rm syn}=I_{\rm syn}-Q_{\rm syn}$ for synchrotron radiation, whereas if $\Psi=0$, $P_{\rm jit}=I_{\rm jit}+Q_{\rm jit}$ for jitter radiation and $P_{\rm syn}=I_{\rm syn}+Q_{\rm syn}$ for synchrotron radiation (see also Eqs. (\ref{synps}) and (\ref{jitps})). This means that the flux is larger in the direction parallel to $\textbf{\emph{s}}_{\rm proj}$. Therefore, the polarization vector \textbf{\emph{e}} is parallel to $\textbf{\emph{s}}_{\rm proj}$.

Let us write $\textbf{\emph{j}}={\rm sin}\theta{\rm cos}\theta_{\rm pos}\hat{e}_N+{\rm sin}\theta{\rm sin}\theta_{\rm pos}\hat{e}_E+{\rm cos}\theta\hat{e}_n$, $\textbf{\emph{s}}={\rm sin}\varphi{\rm cos}\varphi_{\rm pos}\hat{e}_N+{\rm sin}\varphi{\rm sin}\varphi_{\rm pos}\hat{e}_E+{\rm cos}\varphi\hat{e}_n$, and $\textbf{\emph{n}}=\hat{e}_n$, where $\hat{e}_E$ is the unit vector on the sky plane pointing to East. Once the polarization angle $\Phi$ is determined by observation, one can obtain $\varphi_{\rm pos}$ in terms of the angle relation $\varphi_{\rm pos}=\Phi$. Furthermore,
provided that $\theta$ and $\varphi$ are fixed, as well as the position angle of the jet $\theta_{\rm pos}$ on the sky plane is known by observation, we can derive the angle $\alpha_{\rm pos}$ and the angle $\vartheta$ between \textbf{\emph{j}} and \textbf{\emph{s}}.

\subsection{Polarization Processes}
\subsubsection{Polarization of Synchrotron Emission}

We present the some main formulae related to the current work, on the basis of \cite{Prosekin16}'s works. The power spectrum of synchrotron radiation of relativistic electrons in the tensor form is expressed as
\begin{equation}
P^{\rm syn}_{ik}=\frac{\sqrt{3}e^2}{4\pi R_{\rm L}}[I_{\rm syn}\delta_{ik}-Q_{\rm syn}(\delta_{ik}-2s_{{\rm proj},i} s_{{\rm proj},k})],
\end{equation}
where, $\delta_{ik}$ is the two-dimensional Kronecker delta perpendicular to the direction \textbf{\emph{n}}. $I_{\rm syn}$ and $Q_{\rm syn}$ correspond to the prefactor to the Stokes parameters $I$ and $Q$, respectively. They are given by

\begin{equation}
I_{\rm syn}(\omega)=\pi \int_{z_{\rm 0}}^{z_{\rm max}}{\rm d}z\int^{\gamma_{\rm max}}_{\gamma_{\rm min}}{\rm d}\gamma\tilde{N}_{\rm \gamma}(\gamma,z)I_{\rm syn}^{\rm mon}(\omega)R_{\rm jet}^2(z), \label{Isyn}
\end{equation}
\begin{equation}
Q_{\rm syn}(\omega)=\pi\int_{z_{\rm 0}}^{z_{\rm max}}{\rm d}z\int^{\gamma_{\rm max}}_{\gamma_{\rm min}}{\rm d}\gamma\tilde{N}_{\rm \gamma}(\gamma,z)Q_{\rm syn}^{\rm mon}(\omega) R_{\rm jet}^2(z). \label{Qsyn}
\end{equation}
Here the functions $I_{\rm syn}^{\rm mon}$ and $Q_{\rm syn}^{\rm mon}$ arise from a monoenergetic electron emission, and are written as (\citealt{Prosekin16})

\begin{equation}
I_{\rm syn}^{\rm mon}(\omega)=\frac{2}{\pi}\int^{\pi/2}_{0}\chi F(\xi/\chi){\rm d}\phi_{B},\label{Isynmon}
\end{equation}
\begin{equation}
Q_{\rm syn}^{\rm mon}(\omega)=\frac{2}{\pi}\int^{\pi/2}_{0}\left [\frac{2(\chi^2-\sigma^2)}{1-\sigma^2}-\chi^2\right]\frac{1}{\chi} G(\xi/\chi){\rm d}\phi_{B},\label{Qsynmon}
\end{equation}
where $\chi=[1-(1-\sigma^2){\rm cos}^2\phi_B]^{1/2}$, $\xi=\omega/\omega_{\rm c}$, $\omega=2\pi\nu$, $\omega_{\rm c}=\frac{3}{2}\gamma^2\omega_{\rm B}$ being the critical frequency of the synchrotron emission, and $\omega_{\rm B}=\frac{eB_{\rm ls}(z)}{mc}=c/R_{\rm L}$  being the cyclotron frequency of electrons. Here, $B_{\rm ls}(z)$ is the large-scale turbulent magnetic field in the jet, which is considered as $B_{\rm ls}(z)=B_{\rm ls,0}z_{\rm 0}/z$, and $B_{\rm ls,0}$ is the magnetic field strength at the starting position $z_{\rm 0}$ of a dissipation region. In Equations (\ref{Isynmon}) and (\ref{Qsynmon}), the functions $F(x)$ and $G(x)$ are expressed as
\begin{equation}
F(x)=x\int^{\infty}_{x}K_{\rm 5/3}(\zeta){\rm d}\zeta,~~~~~ G(x)=xK_{\rm 2/3}(x),
\end{equation}
where $K_{\rm 5/3}(\zeta)$ and $K_{\rm 2/3}(x)$ are the modified Bessel functions. The power spectrum in some particular direction \textbf{\emph{e}}, i.e., the position of electric vector in the plane perpendicular to \textbf{\emph{n}}, is written as
\begin{equation}
P_{\rm syn}=P^{\rm syn}_{ik}e_ie_k=\frac{\sqrt{3}e^2}{4\pi R_{\rm L}}[I_{\rm syn}-Q_{\rm syn}(1-2{\rm cos}^2\Psi)]. \label{synps}
\end{equation}
Using Equations (\ref{Isyn}) and (\ref{Qsyn}), the degree of polarization of synchrotron emission of electron population is written as
\begin{equation}
\Pi_{\rm syn}(\omega)=\frac{Q_{\rm syn}(\omega)}{I_{\rm syn}(\omega)}.
\end{equation}
Integrating the power spectrum $P_{\rm syn}$ over the frequency $\omega$, we can obtain the polarization degree of the total synchrotron radiation to be
\begin{equation}
\Pi_{\rm syn}=\frac{\int Q_{\rm syn}(\omega){\rm d}\omega}{\int I_{\rm syn}(\omega){\rm d}\omega}=\frac{3}{4}\left(\frac{1-\sigma^2}{1+\sigma^2}\right). \label{totalIIsyn}
\end{equation}
As shown below, jitter radiation has the same polarization degree for its total radiation intensity.

\subsubsection{Polarization of Jitter Emission}
The power spectrum of jitter radiation of relativistic electrons in the tensor form is written as
\begin{equation}
P^{\rm jit}_{ik}=\frac{e^4\left<B_{\rm ss}^2\right>}{m^2c^4}[I_{\rm jit}\delta_{ik}-Q_{\rm jit}(\delta_{ik}-2s_{{\rm proj},i} s_{{\rm proj},k})],
\end{equation}
where $B_{\rm ss}$ is the strength of small-scale turbulent magnetic field in the jet. The paper assumes that $B_{\rm ss}$ has the same spatial distribution along the jet as the large-scale magnetic field, i.e., $B_{\rm ss}=\varsigma B_{\rm ls}$, where, $\varsigma$ is the ratio factor. The functions $I_{\rm jit}$ and $Q_{\rm jit}$ correspond to the prefactor of the Stokes parameters $I$ and $Q$, respectively. They are determined by
\begin{equation}
I_{\rm jit}(\omega)=\pi\int_{z_{\rm 0}}^{z_{\rm max}}{\rm d}z\int^{\gamma_{\rm max}}_{\gamma_{\rm min}}{\rm d}\gamma\tilde{N}_{\rm \gamma}(\gamma,z)I_{\rm jit}^{\rm mon}(\omega) R_{\rm jet}^2(z), \label{Ijitter}
\end{equation}
\begin{equation}
Q_{\rm jit}(\omega)=\pi \int_{z_{\rm 0}}^{z_{\rm max}}{\rm d}z\int^{\gamma_{\rm max}}_{\gamma_{\rm min}}{\rm d}\gamma\tilde{N}_{\rm \gamma}(\gamma,z)Q_{\rm jit}^{\rm mon}(\omega)R_{\rm jet}^2(z), \label{Qjitter}
\end{equation}
where the functions $I_{\rm jit}^{\rm mon}$ and $Q_{\rm jit}^{\rm mon}$ are given by (see \citealt{Prosekin16})
\begin{equation}
I_{\rm jit}^{\rm mon}(\omega)=\frac{1}{4\pi^2}\left(\frac{\omega}{2\gamma^2c}\right)^2\int^1_0 \frac{1}{\eta^3}\varpi(\frac{\omega}{2\gamma^2c\eta})\digamma_1(\eta,\sigma)
{\rm d}\eta,\label{Ijittermon}
\end{equation}
\begin{equation}
Q_{\rm jit}^{\rm mon}(\omega)=\frac{1}{4\pi^2}\left(\frac{\omega}{2\gamma^2c}\right)^2\int^1_0 \frac{1}{\eta^3}\varpi(\frac{\omega}{2\gamma^2c\eta})\digamma_2(\eta,\sigma)
{\rm d}\eta.\label{Qjittermon}
\end{equation}
Here, the functions $\digamma_1(\eta,\sigma)$ and $\digamma_2(\eta,\sigma)$ are
\begin{equation}
\begin{aligned}
\digamma_1(\eta,\sigma)=\left[\frac{2\sigma}{3}+(2-\frac{1}{\sigma})\eta^2+\left(\frac{1}{3\sigma^2}-1\right)
\eta^3-\eta(1+{\rm ln}\sigma) \right]\\
\Theta(\sigma-\eta)+\eta(2\eta-\eta^2-1-{\rm ln}\eta)\Theta(\eta-\sigma),
\end{aligned}
\end{equation}

\begin{equation}
\begin{aligned}
\digamma_2(\eta,\sigma)=\frac{1}{3}\left(\frac{1-\sigma}{1+\sigma} \right) \left[\frac{(1+\sigma)^2\eta^3}{2\sigma^2} +\sigma\right]\Theta(\sigma-\eta)+\\ \frac{1}{1-\sigma^2}\left[\frac{1+\sigma^2}{2}(1-\eta^2)\eta-\frac{2\sigma^2}{3}(1-\eta^3) \right]\Theta(\eta-\sigma),
\end{aligned}
\end{equation}
where $\Theta$ is the Heaviside step function. The radiation intensity in the direction \textbf{\emph{e}} is expressed as
\begin{equation}
P_{\rm jit}=P^{\rm jit}_{ik}e_ie_k=\frac{e^4\left<B_{\rm ss}^2\right>}{m^2c^4}[I_{\rm jit}-Q_{\rm jit}(1-2{\rm cos}^2\Psi)]. \label{jitps}
\end{equation}
Therefore, the degree of polarization of jitter radiation per angular frequency is given by
\begin{equation}
\Pi_{\rm jit}(\omega)=\frac{Q_{\rm jit}(\omega)}{I_{\rm jit}(\omega)}.
\end{equation}
After integrating the power spectrum $P_{\rm jit}$ over the angular frequency $\omega$, we can immediately obtain the polarization degree of the total jitter radiation
\begin{equation}
\Pi_{\rm jit}=\frac{\int Q_{\rm jit}(\omega){\rm d}\omega}{\int I_{\rm jit}(\omega){\rm d}\omega}=\frac{3}{4}\left(\frac{1-\sigma^2}{1+\sigma^2}\right). \label{totalIIjit}
\end{equation}
Thus, it can be seen that the polarization degree is consistent with that of the synchrotron radiation. The function $\varpi$ included in the integrand of Equations (\ref{Ijittermon}) and (\ref{Qjittermon}) describes the distribution of turbulence, and is written as
\begin{equation}
\varpi(\kappa)=\frac{K_\alpha\lambda^3}{(1+\lambda^2\kappa^2)^{1+\alpha/2}}. \label{TurEq}
\end{equation}
The derivation of the normalization coefficient $K_\alpha$ satisfies the non-divergent condition of the magnetic field. It is expressed as
\begin{equation}
K_\alpha=8\pi^{\frac{3}{2}}\frac{\Gamma(1+\alpha/2)}{\Gamma[(\alpha-1)/2]},
\end{equation}
where $\Gamma$ is the gamma function.

As seen in Equation (\ref{TurEq}), jitter radiation depends on the correlation scale, $\lambda$, of a turbulent magnetic field and the spectral index of turbulence $\alpha$. The famous Kolmogorov spectrum presents the index $\alpha=5/3$, describing the turbulence in hydrodynamics. However, magnetohydrodynamical (MHD) turbulence is more complex than the hydrodynamical one. In the case of incompressible turbulence, \cite{Goldreich95} claimed that spectral index is compatible with hydrodynamical scenario.
Simulations of compressible MHD turbulence showed that turbulent fluctuations are divided into three types (\citealt{Cho02}): (1) Alfv\'{e}n wave, which has a Kolmogorov-type spectrum, $\alpha=5/3$;  (2) slow mode, which follows the spectrum of the Alfv\'{e}n mode, $\alpha=5/3$; (3) fast mode, which corresponds to the spectrum of acoustic turbulence with the index $\alpha=3/2$. In the case of shock environment, turbulence spectrum is expected to give rise to a steeper spectral index $\alpha \sim 2$ (\citealt{Padoan09,Chep10}).

\section{Numerical Results}

\begin{figure} \begin{center}
        \includegraphics[width=90mm,height=80mm,bb=19 220 510 600]{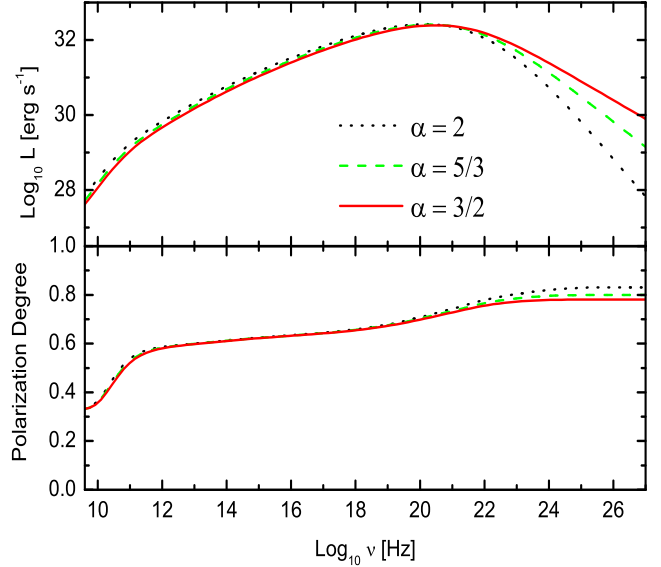}
\caption[ ]{The SEDs (upper panel ) and the degree of polarization (lower panel) of jitter radiation as a function of the frequency at $\varphi=80^\circ$, for the different turbulence spectral indices $\alpha=$2, 5/3, and 3/2.  } \label{fig:JitterPS}
    \end{center}
\end{figure}

In this section, we explore the properties of polarization and SEDs by using some typical parameters for a general high-mass X-ray binary, i.e., microquasar. The typical values of this system are given as follows: a black hole mass
of $M_{\rm BH}=20M_{\odot}$, a distance of $d=2$ kpc, an effective surface temperature of the stellar companion of $T=10^4$ K, a radius of the companion of $R_{\rm co}=20R_{\odot}$, an orbital radius of binary system of $R_{\rm orb}\sim 10^{12}$ cm, an opening angle of the jet of $5$ degrees, and a bulk Lorentz factor of the jet of $\Gamma_{\rm j}=1.5$.

Regarding numerical procedures to solve Equation (\ref{dNdz2}), interested readers are referred to \cite{Zhang14}, in which we have studied in detail the evolution of relativistic electrons along the jet. The model parameters are: the starting point of the dissipation region $z_{\rm 0}$; the terminal position of the dissipation region $z_{\rm max}$; the spectral index of relativistic electrons $p$; the break energy of electrons $\gamma_{\rm cut}$; the minimum Lorentz factor of electrons $\gamma_{\rm min}$; the characteristic length of turbulence $\lambda$; the angle $\varphi$ between the normal to the slab and the line of sight, and the magnetic field strength $B_{\rm ls,0}$. These parameter values are listed in Table \ref{table:cases} for each figure throughout the paper. In the studies of this section, the viewing angle of an observer is set as  $30^\circ$.

\begin{figure} \begin{center}
        \includegraphics[width=90mm,height=80mm]{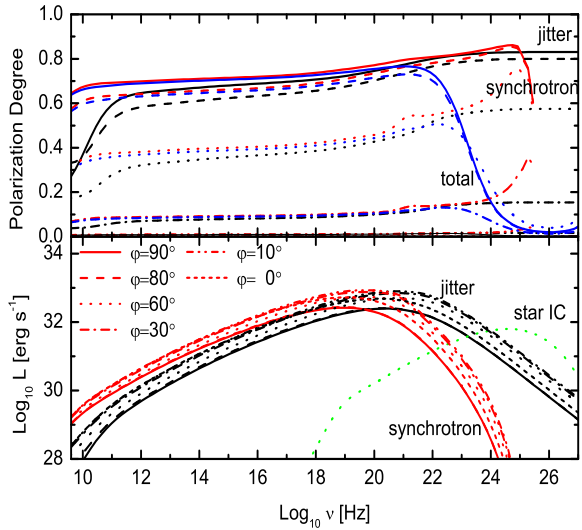}
\caption[ ]{ The degrees of polarization (upper panel) and SEDs (lower panel) of both the synchrotron and jitter radiation as a function of the frequency, with a ratio factor, $\varsigma=1$, of turbulent magnetic field strengths, and the turbulence index of $\alpha=5/3$. Different curves indicate individual components corresponding to the angle values $\varphi$ as labeled in the legend. The degree of polarization of the total radiation is shown in the upper panel. The SED of Comptonization of non-polarized soft photons from companion star is plotted in the lower panel.  } \label{fig:angles}
    \end{center}
\end{figure}

We first explore how magnetic turbulence influences on the properties of polarization and SEDs of jitter radiation. For our purposes, the ratio factor $\varsigma$ is set as 1, and the angle $\varphi$ as $80^\circ$. In Figure \ref{fig:JitterPS}, we plot the SEDs and polarization degree of the jitter radiation, with $\alpha=2$, 5/3, and 3/2.  At low frequencies, i.e., below the peak frequency, the change of turbulence index almost does not affect the SEDs and polarization degree of the jitter radiation, and the SED shape is associated with the spectral index of emitting relativistic electrons, which is similar to spectral behaviors of synchrotron emission. Above the peak frequency, the turbulence index determines the shape of SEDs that becomes steeper with increasing the index $\alpha$. As seen in the lower panel, a large value of turbulence index gives rise to a high polarization degree. The large value of the turbulence index implies that more turbulence energies are concentrated at a larger-scale region of turbulence (corresponding to small wavenumber regime). Therefore, the turbulent magnetic field with a large-scale configuration results in a higher polarized radiation at high frequency bands.

It is noticed that the polarization degree first increases sharply with increasing frequency at radio wavebands, and tends to be slow in the $10^{12}$ - $10^{19}$ Hz range, then becomes faster near the peak frequency ($\sim10^{21}$ Hz). After the peak frequency, the distribution of polarization degree tends to form a plateau, i.e., close to a constant polarization degree. In our calculation, we also test the influence of different correlation lengths of small-scale magnetic fields on both SEDs and polarization of jitter radiation. The results demonstrate that the peak frequency of jitter radiation shifts to higher frequencies with decreasing the value $\lambda$. A change of distributions of polarization degree is related in the same way to jitter SEDs as those of Figure \ref{fig:JitterPS}.

We now study how the angle between the line of sight and the normal to the slab of chaotic magnetic fields influences on the polarization degree and SEDs of both jitter and synchrotron emissions. Here, we first fix $\alpha=5/3$ and $\varsigma=1$, then change the angle values in order to observe their behaviors. We in Figure \ref{fig:angles} plot the degree of polarization (upper panel) and SEDs (lower panel) of both the synchrotron and jitter radiation as a function of the frequency. Meanwhile, the non-polarized inverse Compton (IC) scattering of the companion star, which is independent of the angle $\varphi$, is also included in the lower panel. It is evident that the degrees of polarization of both jitter and synchrotron emissions decrease from $\sim86\%$ to $0\%$, with decreasing angle $\varphi$, that is, from $90^\circ$ (observed edge-on) to $0^\circ$ (observed face-on), respectively. Similar to jitter case (see the lower panel of Figure \ref{fig:JitterPS}), synchrotron polarization demonstrates also a sharp increase at radio frequencies, but this increase appears at the lower radio bands. In general, these polarization distributions present plateau features at low frequencies, then increase slowly near peak frequency. As for the same angle $\varphi$, the polarized synchrotron emission has a slight larger polarization degree than jitter radiation. The distribution of the degree of polarization of the total emissions including non-polarized IC process presents an exponent-like form at high frequencies. It should be mentioned that a sharp cutoff of synchrotron polarization degree at high-frequency limit is a numerical artifact because the relation $\Pi=Q_{\rm syn}/I_{\rm syn}$ remains finite.

As shown in the lower panel of Figure \ref{fig:angles}, both jitter and synchrotron emissions have significantly different spectral shapes at high-frequency limit. The synchrotron emission decreases exponentially over a break frequency $\nu_{\rm cut}=3eB_{\rm ls}\gamma_{\rm cut}^2/4\pi m_{e}c$, whereas the jitter radiation spectrum beyond the break frequency $\nu_{\rm cut}R_{\rm L}/\lambda$ presents a long power-law tail, which depends on the turbulence index $\alpha$ but is independent of a distribution of emitting electrons (see also Figure \ref{fig:JitterPS}). This unique feature of jitter radiation implies direct, model-independent information about the properties of turbulence spectrum. However, synchrotron emission can provide information on energy spectral shapes of emitting relativistic electrons, which is associated with different acceleration mechanisms (see also \citealt{Kelner13}, for more discussions). With decreasing the angle $\varphi$, intensities of both jitter and synchrotron radiation increase, which are anti-correlated with their linear polarization degrees.

\begin{figure} \begin{center}
        \includegraphics[width=90mm,height=80mm]{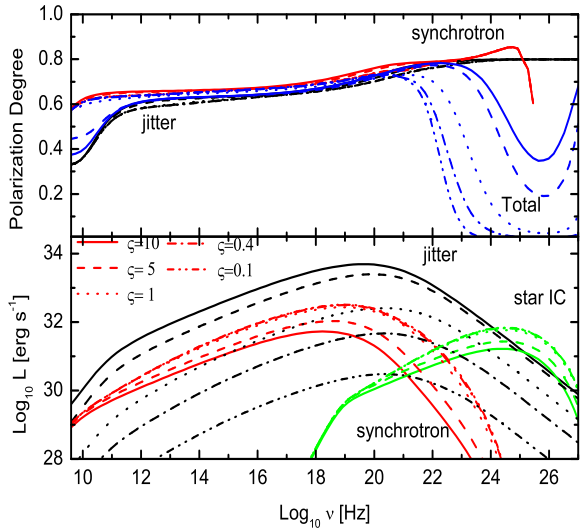}
\caption[ ]{The degrees of polarization (upper panel) and SEDs (lower panel) of both synchrotron and jitter radiation as a function of the frequency, with $\varphi=80^\circ$ and $\alpha=5/3$, for different values of the ratio factor $\varsigma$ labeled in the legend. The degree of polarization of the total radiation is shown in the upper panel. The SEDs of the non-polarized IC scattering are also shown in the lower panel.  } \label{fig:Bratio}
    \end{center}
\end{figure}

Below, we investigate how relative strengths between large- and small-scale turbulent magnetic fields influence on the polarized jitter and synchrotron emissions as well as IC scattering spectrum of the surrounding companion. Here, we fix $\varphi=80^\circ$ and $\alpha=5/3$, but change the ratio factor $\varsigma$. Figure \ref{fig:Bratio} presents the resulting polarization distributions (upper panel) and SEDs (lower panel). The SEDs of Comptonization of non-polarized soft photons from companion star are plotted in the lower panel, which depends on the relative strength of turbulent magnetic fields. It is easy to understand this phenomenon because the evolution of injected electrons is related to jitter, synchrotron, IC losses by Equation (\ref{dNdz2}). The variations of the factor $\varsigma$ would result in the change of spectral distributions of emitting relativistic electrons, producing different photon spectral features. As shown in the lower panel, when small-scale magnetic field dominates, i.e., $\varsigma>1$, jitter radiation dominates emission output. When $\varsigma\leq 1$, synchrotron components dominate below the characteristic frequency of the jitter radiation (about $10^{21}$ Hz). At high-frequency limit, jitter radiation is dominant component due to its typical spectral feature relevant to turbulence nature (see the previous discussions related to Figure \ref{fig:angles}). With decreasing the value $\varsigma$, two component losses including both jitter and synchrotron decrease compared to IC losses of star, the reason why star IC fluxes increase. As seen in the upper panel, the change of the ratio factor $\varsigma$ almost does not influence on the degrees of polarization from individual jitter and synchrotron emissions. But, due to the presence of the non-polarized IC component, the degree of polarization of the total emissions demonstrates a trough feature at high frequency bands.

In the above studies, the viewing angle, that is, the angle between the line of sight and the jet axis is set as $\theta=30^\circ$, which gives a constant Doppler factor via $D=1/\Gamma_{\rm j}(1-\beta_{\rm j}\rm {cos}\theta)$. For relativistic motion of the jet, the variation of the angle $\theta$ would only effect on the amplitude of radiation fluxes rather than the SED shape. Meanwhile, the change of the viewing angle does not result in the change of the polarization degree due to both the jitter and synchrotron emissions, for the same value of the angle $\varphi$. However, an increase of the viewing angle $\theta$ would lead to the decrease in the angle $\vartheta$ between the jet axis \textbf{\emph{j}} and the normal to the slab of magnetic fields \textbf{\emph{s}} via the relation ${\rm cos\vartheta}=\textbf{\emph{j}}\cdot\textbf{\emph{s}}$.

Similar to a common procedure adopted in studies of extragalactic AGN and Galactic jets, the related numerical calculations of radiation processes in this study are carried out in the co-moving frame of the jet, then the results are transformed to the reference frame of the observer to reproduce observations. Alternatively, provided that the calculation is performed in the observer's reference frame, and some structures of the slabs in the jet reference frame can be determined, one should recalculate all the directions of slabs taking into account relativistic aberration, via the transformation relation, ${\rm tan}\varphi'={\rm sin}\varphi/\Gamma_{\rm j}({\rm sin}\varphi-\beta_{\rm j})$.

\section{Application to Cygnus X--1}
\label{Appl}

\begin{figure*}[]
\centerline{\includegraphics[width=80mm,height=70mm]{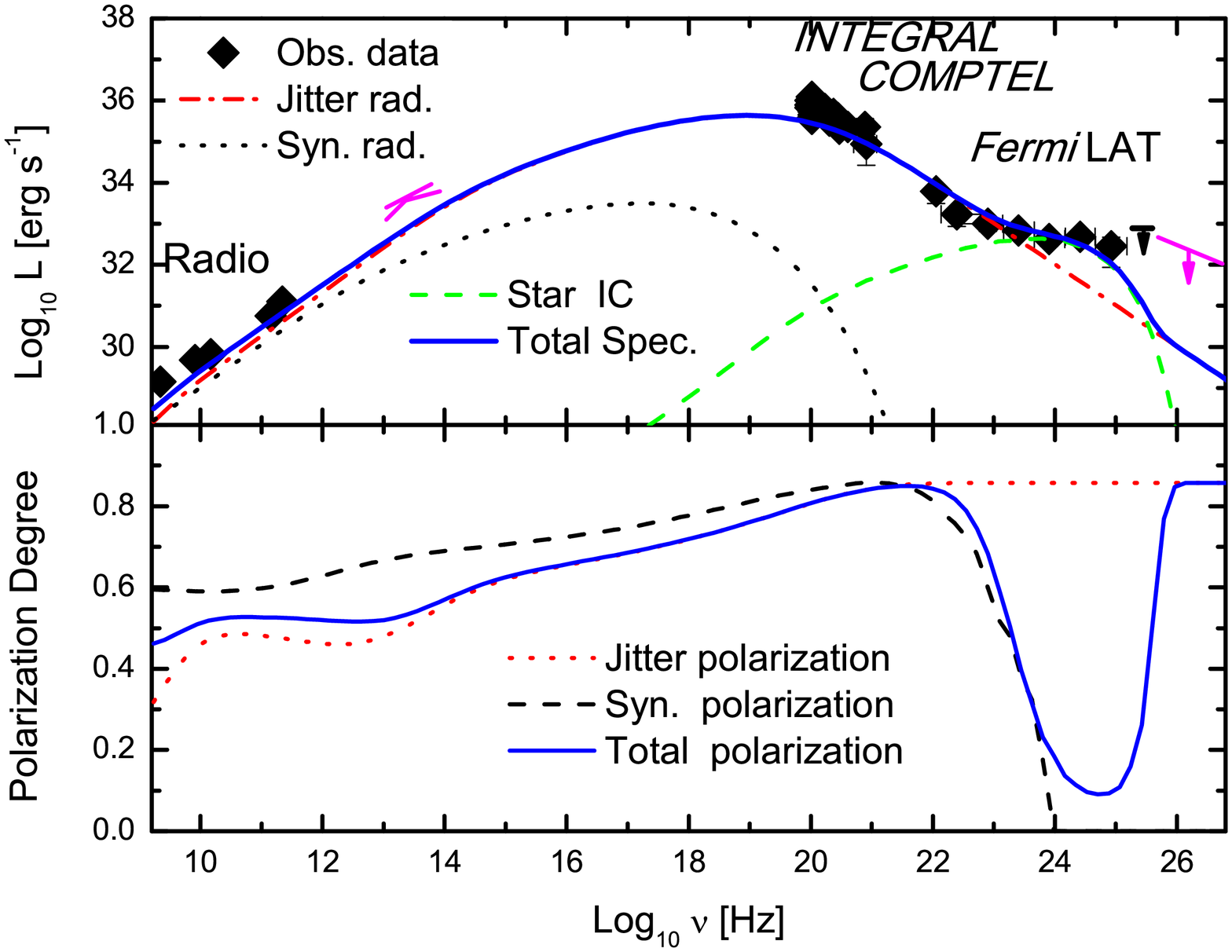}
\includegraphics[width=80mm,height=70mm]{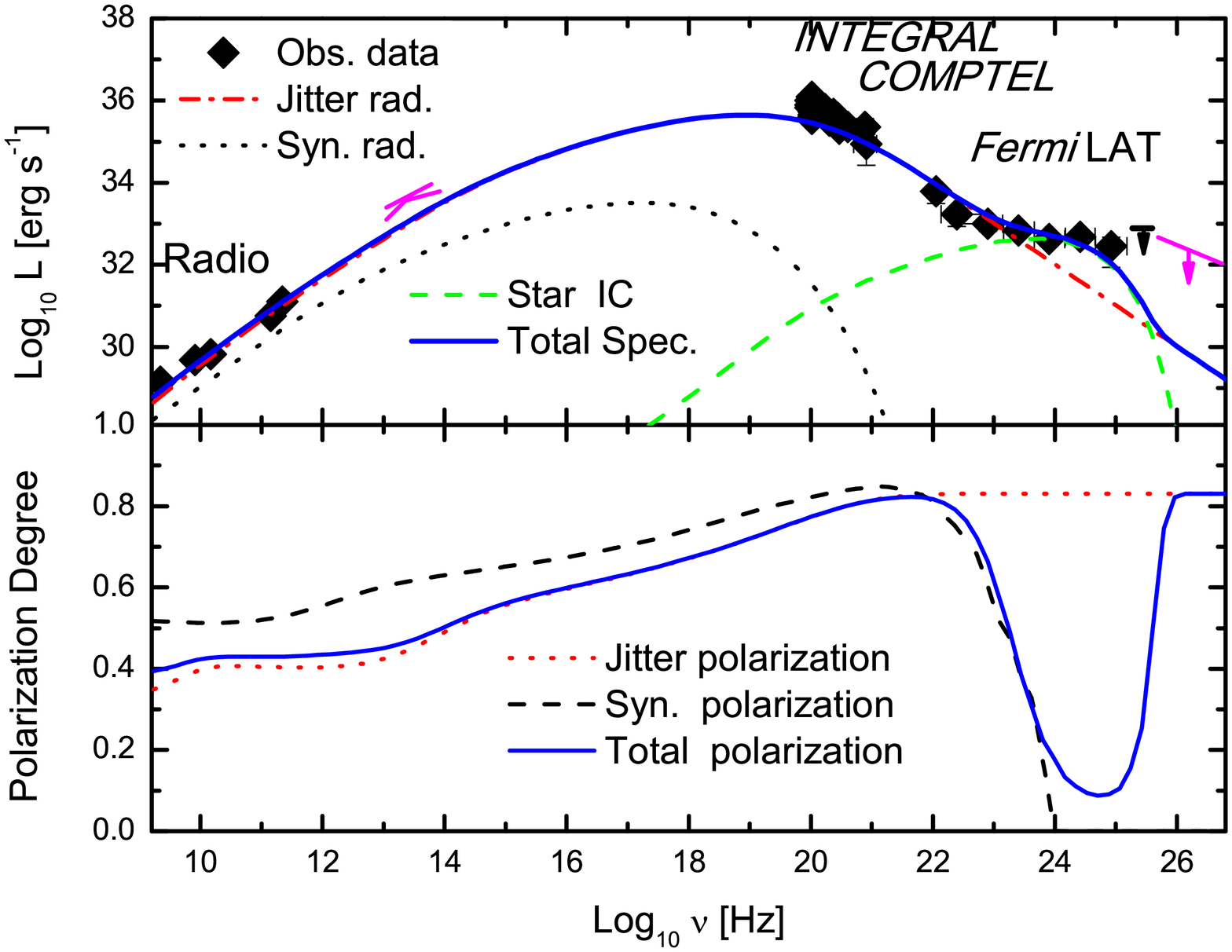}}
\centerline{\includegraphics[width=80mm,height=70mm]{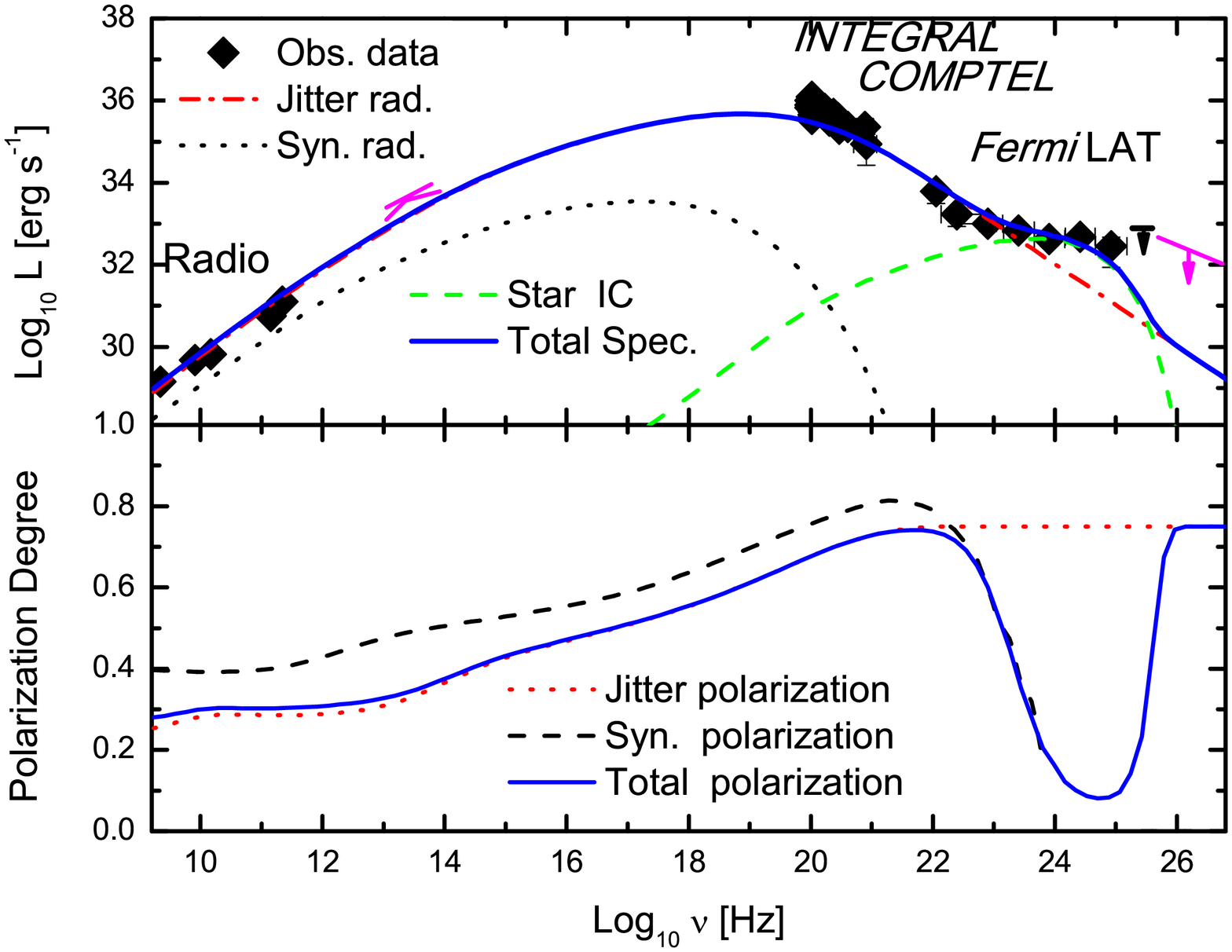}
\includegraphics[width=80mm,height=70mm]{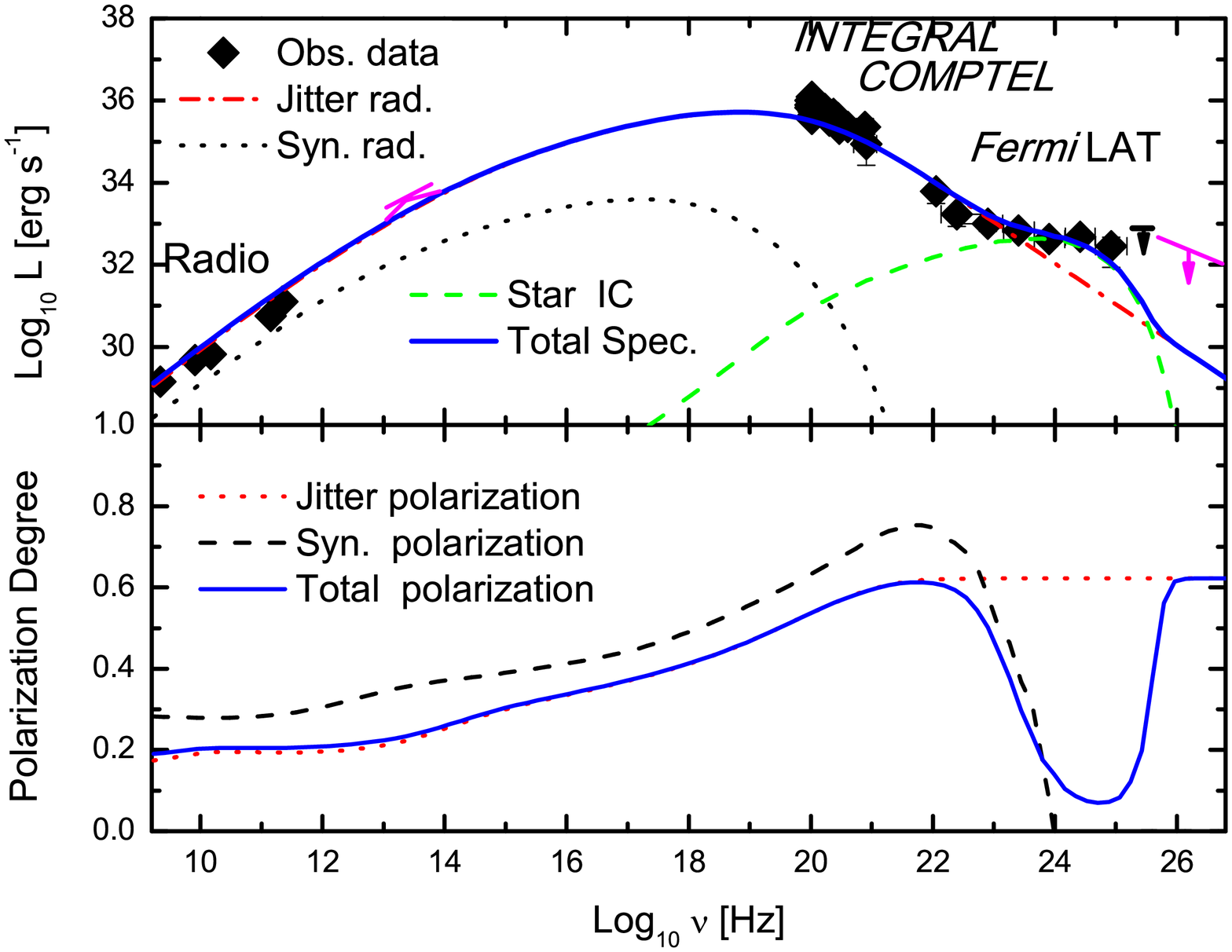}}
\caption{The fitting of multiband observations and the degree of polarization for Cygnus X--1, in the case of dominant small-scale turbulent magnetic field. The plotted observations are radio data from \cite{Fender00}, hard X-ray data points from \cite{Zdziarski12} by INTEGRAL, soft $\gamma$-ray data from \cite{McConnell11} by COMPTEL, Fermi LAT observations from \cite{Zdziarski16}. The lower limit of IR emission observed, which has a high flux and is usually explained as the blackbody radiation of the stellar companion, is marked as `K' glyph. The upper limits of MAGIC is marked approximately on the figure according to \cite{Albert07}. The angles between the line of sight and the normal to the slab of turbulent magnetic fields are $\varphi=90^\circ$ (left upper panel), $80^\circ$ (right upper), $70^\circ$ (left lower), and $60^\circ$ (right lower), respectively. Different curves indicate individual and total polarization and radiation spectral components as labeled in the legend.
}  \label{figs:fittingtail}
\end{figure*}

Cygnus X-1 is a high-mass X-ray binary, in which the central black hole identified is accreting matter from the stellar companion (\citealt{Orosz11}).
The mass of the former is measured to be $16M_{\odot}$, but the mass of the latter remains relatively uncertain and here is considered as $27M_{\odot}$ (e.g., \citealt{Zdziarski16}). The system is located at a distance of 1.86 kpc (\citealt{Reid11,Xiang11}), which has a separation distance, $3.2\times10^{12}$ cm, between two components, with an orbital period of 5.6 days. The parameters related to the companion star are its effective surface temperature of $\sim 2.8\times10^{4}$ K, and its radius of $16R_{\odot}$, which immediately gives the monochromatic luminosity $\sim10^{38}$ erg $\rm s^{-1}$, according to the Stefan-Boltzmann law. The parameters related to the jet are the bulk Lorentz factor of $\Gamma_{\rm j}=1.25$ (\citealt{Zdziarski14,Zhang14,Zdziarski16}), and the half-opening angle of the jet of $0.5^\circ$ (\citealt{Stirling01,Zdziarski14,Zdziarski16}). In addition, an inclination of the normal to the orbital plane with respect to the line of sight is $29^\circ$ (\citealt{Orosz11,Zilkowski14}).

As discussed in the introduction section, the origin of MeV tail emission and polarization is still an open question. Our works focus on understanding the properties of its radiation and polarization in a turbulent environment mixing large- and small-scale magnetic fields. Based on the relative strength between large-scale turbulent magnetic field and small-scale one, we explore three scenarios: (1) the case of a dominant small-scale turbulent environment ($\varsigma>1$), jitter radiation contributing to radio, MeV tail and \emph{Fermi} LAT (to low frequency bands, $<10^{23}$ Hz) observations; (2) an equipartition scenario of the strength between two turbulent fields ($\varsigma=1$), synchrotron emission reproducing radio flux, and jitter radiation dominating \emph{Fermi} LAT ($<10^{23}$ Hz) observations; (3) the case of dominant large-scale fields ($\varsigma<1$), synchrotron radiation emitting radio flux, as well as both jitter and synchrotron emissions contributing to \emph{Fermi} LAT ($<10^{23}$ Hz) observations.

\begin{deluxetable*}{cccccccccccc}
\tabletypesize{}
\tablecaption{The Model Parameters Used in the Study.}
\tablewidth{0pt}
\tablehead{
\colhead{Case} & \colhead{$\varsigma$} & \colhead{$\lambda[R_{\rm L}]$} &
\colhead{$\alpha$} & \colhead{$z_{\rm 0}[R_{\rm orb}]$} & \colhead{$z_{\rm max}[R_{\rm orb}]$} & \colhead{$B_{\rm ls,0}[G]$} & \colhead{$\eta_{\rm rel}$} & \colhead{$p$} & \colhead{$\gamma_{\rm min}$} & \colhead{$\gamma_{\rm cut}$} & \colhead{$\varphi$}}
\startdata
Fig. 2 & 1 & 0.1 & X  & 0.01 & 50 & 100 & 0.3 & 2 & 50 & $5\times10^6$ & $80^\circ$ \\
Fig. 3 & 1 & 0.1 & $5/3$  & 0.01 & 50 & 100 & 0.3 & 2 & 50 & $5\times10^6$ & X \\
Fig. 4 & X & 0.1 & $5/3$  & 0.01 & 50 & 100 & 0.3 & 2 & 50 & $5\times10^6$ & $80^\circ$ \\
Fig. 5 & 12 & 0.03 & $2$  & 0.07 & 1 & 420 & 0.06 & 1 & 1 & $6\times10^4$ & X \\
Fig. 6 & 1 & 0.3 & $5/3$  & 0.05 & 15 & 60 & 0.3 & 2 & 60 & $8\times10^6$ & X
\enddata
\tablenotetext{*}{Note. Symbol indicating $\varsigma$: ratio of small- to large-scale magnetic field; $\lambda$: coherence length of magnetic field; $\alpha$: turbulence spectral index; $z_{\rm 0}$: onset of dissipation; $z_{\rm max}$: end of dissipation; $B_{\rm ls,0}$: magnetic field strength; $\eta_{\rm rel}$: transform factor of electrons; $p$: electron spectral index; $\gamma_{\rm min}$: electron minimum energy; $\gamma_{\rm cut}$: electron break energy; $R_{\rm L}$: non-relativistic Larmor radius; $R_{\rm orb}$: orbital radius of binary system; $\varphi$: the angle between the normal to the slab and the line of sight; X: indicating the change of corresponding parameters.}
\label{table:cases}
\end{deluxetable*}

\begin{figure*}[]
\centerline{\includegraphics[width=80mm,height=70mm]{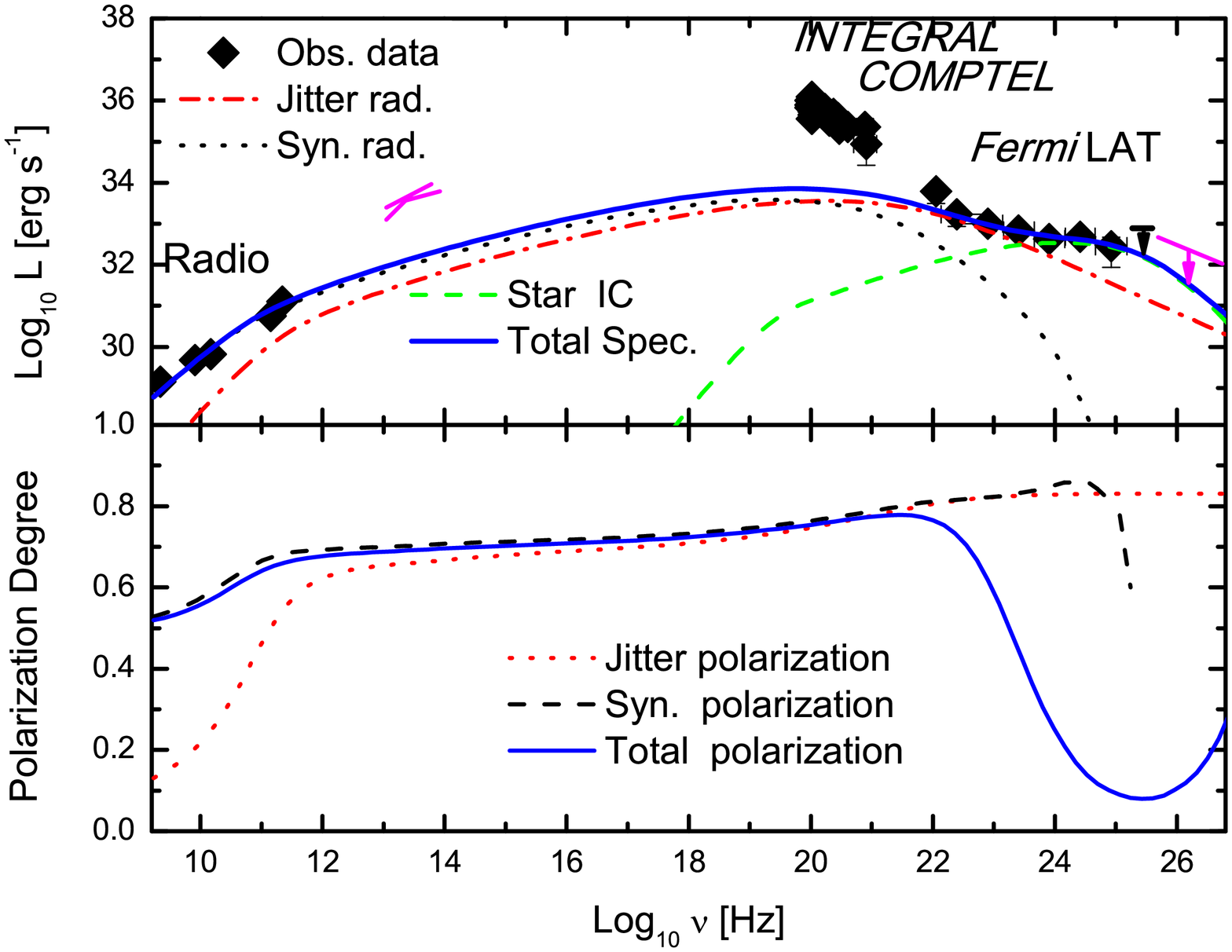} \includegraphics[width=80mm,height=70mm]{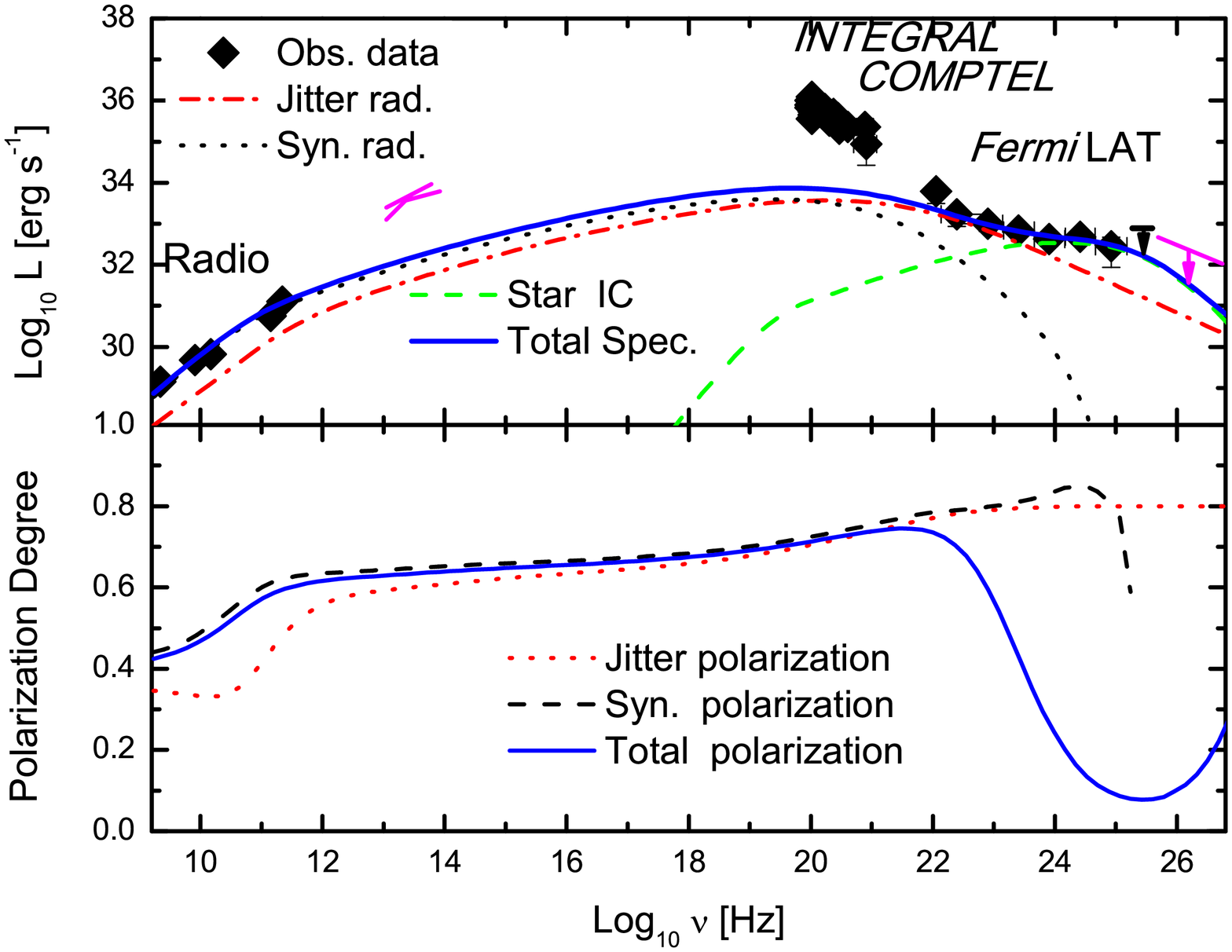}}
\centerline{\includegraphics[width=80mm,height=70mm]{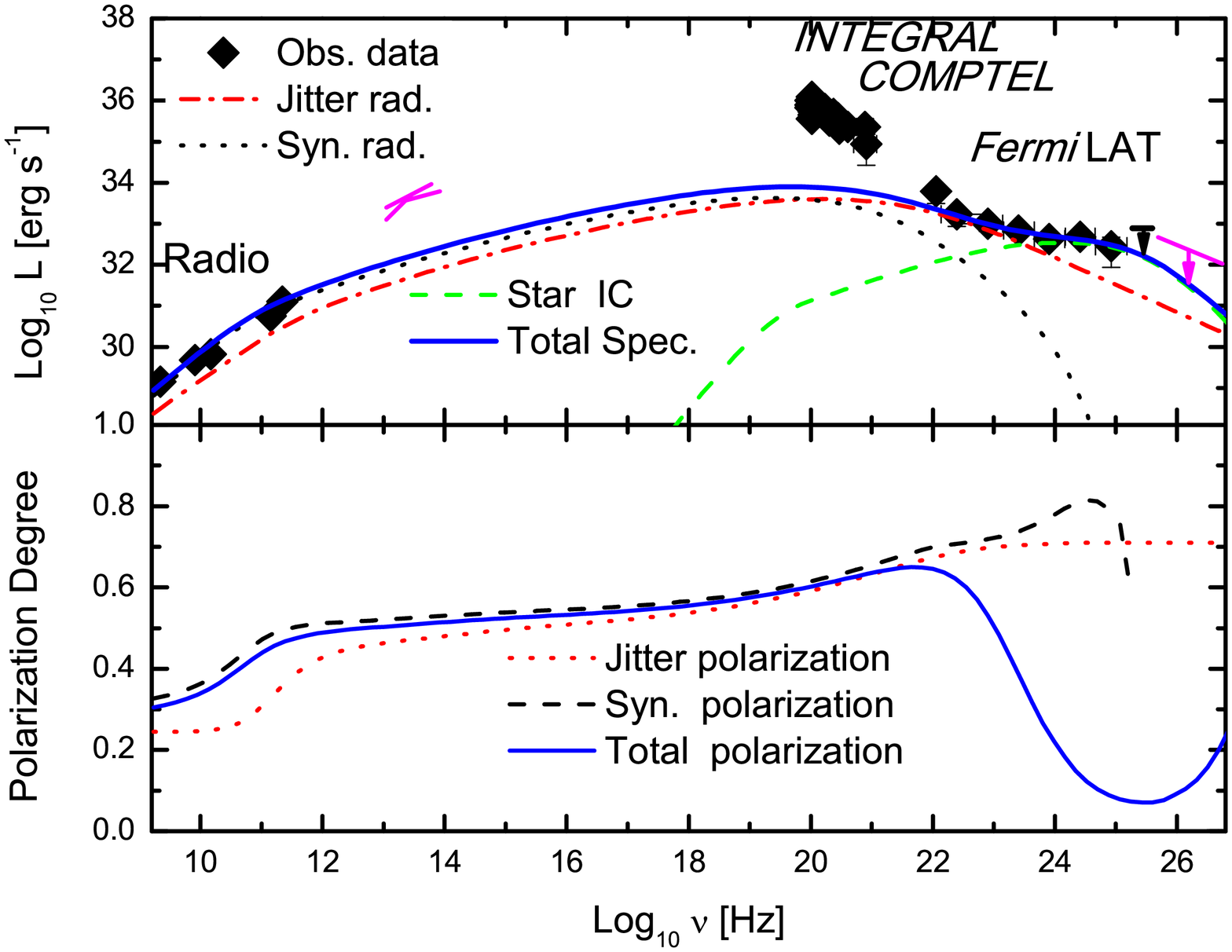} \includegraphics[width=80mm,height=70mm]{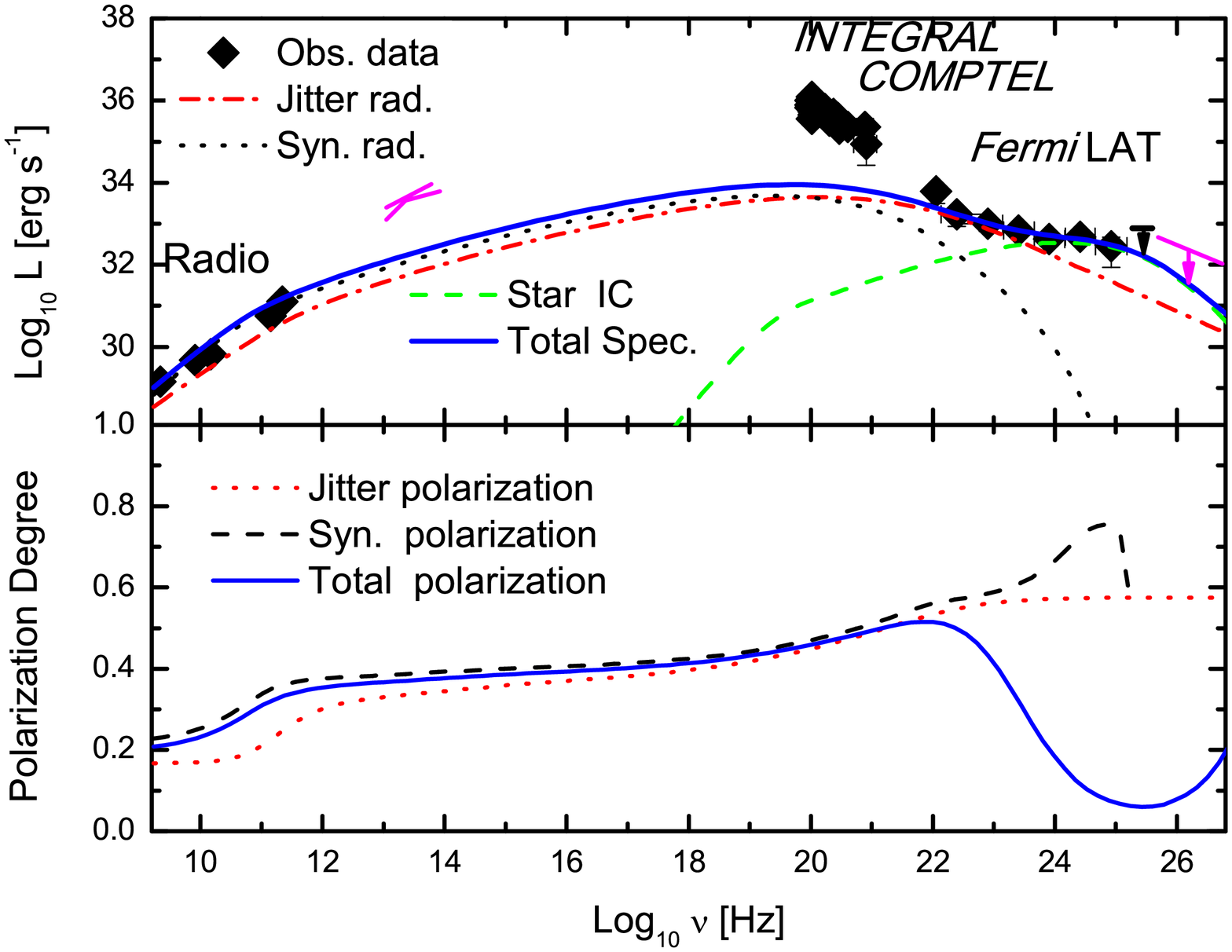}}
\caption{The fitting of multiband observations and the degree of polarization for Cygnus X--1, in the case of equipartition turbulent magnetic field. The angles between the line of sight and the normal to the slab of turbulent magnetic fields are $\varphi=90^\circ$ (left upper panel), $80^\circ$ (right upper), $70^\circ$ (left lower), and $60^\circ$ (right lower), respectively. The other components plotted are the same as those of Figure \ref{figs:fittingtail}. }  \label{figs:fittingGeVjitter}
\end{figure*}

Under the condition of dominant small-scale magnetic fields, we in Figure \ref{figs:fittingtail} present the fitting of multiband observations and the degree of polarization for Cygnus X--1. In this case, our main purpose is to study the properties of MeV tail emission and its polarization. The same parameters used in each panel are listed in Table \ref{table:cases}. In order to obtain a self-consistent result, we in this set of fitting change only the angles $\varphi$ between the line of sight and the normal to the plane of turbulent magnetic fields, which are $\varphi=90^\circ$ (left upper panel), $80^\circ$ (right upper), $70^\circ$ (left lower), and $60^\circ$ (right lower), respectively. It is seen from the figure that jitter radiation reproduces observations at radio, MeV tail, and $\gamma$-ray (about $< 10^{23}$ Hz) bands. The synchrotron component in each panel provides a low, negligible flux throughout the wide waveband, except it has a slight flux contribution at radio bands, for the case of $\varphi=90^\circ$ (see upper-left panel). IC processes of the companion can well explain \emph{Fermi} LAT observations above $10^{23}$ Hz. Take notice that the total emission fluxes at radio bands become stronger with decreasing the value $\varphi$, because a smaller observational angle results in higher jitter and synchrotron emission intensities (see also Figure \ref{fig:angles}).

As shown in each panel of Figure \ref{figs:fittingtail}, synchrotron emission presents an exponent-like cutoff appearing at lower frequencies than that of Figure \ref{fig:angles}; this is because that in our fitting the low break energy $\gamma_{\rm cut}=6\times10^4$ is required. The synchrotron process provides a higher polarization degree than the polarized jitter radiation, but due to low synchrotron emission fluxes it is not promising to detect such a synchrotron polarization signal. As expected, the jitter radiation at MeV tail can provide a high polarization degree. The integrated polarization degrees at soft $\gamma$-ray bands (MeV tail: 0.4--2 MeV) are $\sim81\%$ for $\varphi=90^\circ$, $\sim78\%$ for $\varphi=80^\circ$, $\sim69\%$ for $\varphi=70^\circ$, and $\sim55\%$ for $\varphi=60^\circ$, respectively. The polarization observations demonstrated that the degree of polarization is 67 $\pm$ 30$\%$ between 400 keV and 2 MeV  (\citealt{Laurent11}), 76 $\pm$ 15 $\%$ at 0.23--0.85 MeV (\citealt{Jourd12}), or 75 $\pm$ 32$\%$ between 0.4 MeV and 2 MeV (\citealt{Rodr15}). Hence, our theoretical works can constrain the angle $\varphi$ up to about $75^\circ$, in which the integrated polarization degrees are $\sim75\%$ at MeV tail (0.4--2 MeV), $\sim79\%$ at \emph{Fermi} LAT energy regime (between $10^{22}$ and $10^{23}$ Hz), and $\sim30\%$ at radio bands (between $2\times10^{9}$ and $2\times10^{11}$ Hz).The position angle value of the Cygnus X--1 jet, which is slightly time dependent, is about $21^\circ$ to $24^\circ$ counted from North to East (\citealt{Stirling01}). Noticing $\varphi=75^\circ$ (the angle between \textbf{\emph{s}} and \textbf{\emph{n}}) and $\theta=29^\circ$ (the angle between \textbf{\emph{j}} and \textbf{\emph{n}}), as well as the observational polarization angle $\Phi\approx40^\circ$, we obtain the angle $\vartheta\approx47^\circ$ between \textbf{\emph{j}} and \textbf{\emph{s}}, and the angle $\alpha_{\rm pos}\approx16^\circ$ between $\textbf{\emph{j}}_{\rm proj}$ and $\textbf{\emph{s}}_{\rm proj}$, on the basis of Section \ref{MGeo}.

Our purpose is not to fit the radio emission or explain the corresponding polarization feature. The radio data we have plotted on the figure is to provide a constraint for an upper limit of theoretical SEDs. These radio emissions are usually considered to originate from the region of large-scale jets. Observationally, the polarization degree in this waveband is less than 10 per cent (\citealt{Stirling01}). In general, it needs to consider self-absorption processes of low frequency radio emissions in this case, which would decrease the degree of polarization predicted in this paper, but this is beyond the scope of the paper.

Now, we consider an equipartition scenario ($\varsigma=1$) of strengths between turbulent magnetic fields. The fitting results of multiband observations of Cygnus X--1 and the corresponding polarization degree are plotted in Figure \ref{figs:fittingGeVjitter}, in which the fitting parameters used in each panel are listed Table \ref{table:cases}. As shown in this figure, the synchrotron emission can reproduce radio observations and the jitter
radiation provides emission fluxes at high-energy bands ($<10^{23}$ Hz). The integrated polarization degrees are $\sim59\%$ at radio band mainly from synchrotron polarization, $\sim76\%$ at \emph{Fermi} energy region from the combination of both jitter and synchrotron polarization radiation, for the case of the angle $\varphi=90^\circ$. Similarly, the integrated polarization degrees are given as follows: $\sim53\%$ at radio bands (between $2\times10^{9}$ and $2\times10^{11}$ Hz) and $\sim71\%$ at \emph{Fermi} energies (between $10^{22}$ and $10^{23}$ Hz) for $\varphi=80^\circ$; $\sim40\%$ at radio and $\sim60\%$ at \emph{Fermi} energies for $\varphi=70^\circ$; as well as $\sim29\%$ at radio and $\sim46\%$ at \emph{Fermi} energies for $\varphi=60^\circ$.

We study here the case of dominant large-scale fields ($\varsigma=0.6$). We first fit multiband observations using the following parameters: $\lambda=0.3R_{\rm L}$, $\alpha=5/3$, $z_{\rm 0}=0.05R_{\rm orb}$, $z_{\rm max}=15R_{\rm orb}$, $B_{\rm ls,0}=60$ G, $\eta_{\rm rel}=0.3$, $p=2$, $\gamma_{\rm min}=60$ and $\gamma_{\rm cut}=10^7$. The fitting results show that the radio fluxes are from the synchrotron emission, and \emph{Fermi} LAT fluxes to low frequency bands ($<10^{23}$ Hz) are a combination of jitter and synchrotron emissions. \emph{Fermi} LAT fluxes to high frequency bands ($>10^{23}$ Hz) are due to non-polarized IC processes of the surrounding companion. We do not include these figures in the paper for a conciseness, but they, generally speaking, are similar to that of Figure \ref{figs:fittingGeVjitter}. The integrated polarization degrees we obtain are $\sim61\%$ at radio band due to synchrotron polarization, $\sim75\%$ at \emph{Fermi} energy region from both jitter and synchrotron polarizations, for the case of the angle $\varphi=90^\circ$. In the same way, we have the degrees of polarization: $\sim53\%$ at radio and $\sim72\%$ at \emph{Fermi} energies ($\leq10^{23}$ Hz) for $\varphi=80^\circ$; $\sim41\%$ at radio and $\sim61\%$ at \emph{Fermi} energies for $\varphi=70^\circ$; as well as $\sim29\%$ at radio and $\sim46\%$ at \emph{Fermi} energies for $\varphi=60^\circ$.

The fitting procedures used above are similar to those presented in \cite{Zhang14}, in which we have studied the origin of multiband emission in a certain region of the jet. Here, we simply express the motivations for the parameter choice. Based on the studies in \cite{Zhang14}, in which the MeV tail emissions are produced inside the binary system and the GeV band emissions are from the distance close to the binary system scales, we thus fix $z_{\rm max}=R_{\rm orb}$ (see Figure \ref{figs:fittingtail}) and $15R_{\rm orb}$ (see Figure \ref{figs:fittingGeVjitter}) for fitting MeV tail plus \emph{Fermi} LAT data and for only \emph{Fermi} LAT data, respectively. Then, we adjust magnetic field strength $B_{\rm ls,0}$ and $z_{\rm min}$ to calculate radiative fluxes due to jitter, synchrotron and IC scattering processes. Noticing competition between jitter, synchrotron and IC radiation losses, we further change $\varsigma$, which is directly associated with competition between jitter and synchrotron emissions, and indirectly influences on the IC spectra ($\varsigma$ in Figure \ref{figs:fittingtail} is adjusted to 12 due to this reason). Besides, it should be noticed that the parameters $z_{\rm max}$ and $\gamma_{\rm min}$ would impact on emission fluxes in the radio frequency band.

From the above fittings, we find that SED fittings exist relatively large degeneracy. For instance, as for the change of the angle $\varphi$, the model can provide a good fitting, but the degree of polarization distinct from SEDs is very sensitive to the angle values. It is necessary for the presence of the dominant small-scale turbulent magnetic fields to explain the highly polarized hard X-ray/soft $\gamma$-ray emissions, i.e., MeV tail polarized radiation. Thus, it can be seen that the study of polarization radiation is a robust method to uncover the origin of multiband emissions, the structure of magnetic fields, and the properties of turbulence.

\section{Conclusions and Discussion}
In this paper, we have studied the properties of polarized radiation from X-ray binaries, by assuming the existence of the turbulent magnetic field environment in the jet. These turbulent fields are composed of large- and small-scale magnetic field structures, which result in a polarized jitter radiation when the correlation length of turbulence is less than the non-relativistic Larmor radius, i.e., $\lambda\ll R_{\rm L}$, or a polarized synchrotron emission when $\lambda>R_{\rm L}$. We calculate numerically the SEDs and the degree of polarization for a general microquasar. The results show that turbulent magnetic field configurations can indeed provide a high polarization degree. Then, the model is applied to study the properties of polarized radiation of Cygnus X--1. Under the constraint of the fitting of multiband observations, our studies demonstrate that the model can explain the high polarization degree at MeV tail and predict the highly polarized properties at high-energy $\gamma$-ray bands ($<10^{23}$ Hz).

The fittings of Cygnus X--1 showed that the dominant small-scale turbulent magnetic field plays a key role for explaining the high degree of polarization at MeV tail. Moreover, the modelling needs a large index of the turbulence, $\alpha=2$, corresponding to the turbulence in a shock environment (\citealt{Padoan09,Chep10}), which is in agreement with a usual expectation that relativistic electrons are accelerated by the shocks in the jet. However,
in order to explain the MeV tail spectral observation and its high polarization degree, it needs to provide a hard spectral index, $p=1$, of the injected relativistic electrons. The acceleration mechanism of particles is an unsolved problem and subjected to the ongoing debate. Generally, it could be shock acceleration, stochastic acceleration or magnetic reconnection \citep[see][for a brief review]{ZAA14}, or shock interaction in a magnetic reconnection site \citep{Lazarian99,de05,Drury12}. The non-relativistic diffusive shock acceleration, that is, a first-order \emph{Fermi} process, is often considered as the most effective acceleration mechanism, which gives the spectral index close to 2, from a theoretical point of view. In the particle-in-cell simulations,
solid evidence shows that the late-time particle spectrum integrated over the whole reconnection region is a hard power law ($<2$) for high magnetization environments (\citealt{Sironi14}). Furthermore, in the case of a high magnetization, the formation of a hard power-law close to 1 of the energetic particle spectrum is obtained by three-dimensional PIC simulations from a relativistic magnetic reconnection (\citealt{Guo14}). Besides, the analytical studies of shock acceleration in a magnetic reconnection site also gives a hard spectral index 1 (\citealt{Drury12}).

In the case of the dominant large-scale turbulent magnetic field (including an equipartition case), it is hard to explain the MeV tail emission by jitter and/or synchrotron radiation. In other words, the small-scale turbulent magnetic field and resultant jitter radiation play an important role for explaining the highly polarized hard X-ray/soft $\gamma$-rays observed by INTEGRAL. As far as we know, the jitter radiation has usually been discussed in the context of the magnetic field amplification in GRBs (\citealt{Medvedev99}, see also \citealt{Mao13} for a recent work) and the generation of microturbulence (e.g., via Weibel instability) in weakly magnetized shocks (\citealt{Spit08,Medvedev11}). The production of jitter radiation in the so-called small-scale turbulent magnetic field requires the condition that the coherence length of the field be much smaller than the nonrelativistic Larmor radius, $\lambda\ll R_{\rm L}$. In the case of microquasar jet environment, the averaged magnetic field in the emitting region is in the range of about $10^3$ to $1\rm\ G$ along the outflow direction, which implies the turbulence scale to be in the range of $1$ to $10^3\rm\ cm$ via the condition, $\lambda\ll1.7\times10^3(m/m_{e})(B/1\rm G)^{-1} \rm \ cm$.

The presence of the large-scale turbulent magnetic fields would produce the synchrotron emission losses, whose degree of the polarization is slightly higher than the polarization degree of the jitter radiation for spectral indices of electron population used in this study. However, the degree of the polarization of the total synchrotron emission is equal to that of the total jitter radiation (see Equations (\ref{totalIIsyn}) and (\ref{totalIIjit})). For the same mean magnetic field, an electron energy loss due to synchrotron or jitter emission is the same in the large- and small-scale turbulent magnetic fields. However, in the realistic microquasar jet environment, the ratio of the jitter radiation intensity to synchrotron emission one (due to electron population) can be expressed as $r=C(\alpha,p')(R_{\rm L}/\lambda)^{(p'-3)/2}$ (\citealt{Kelner13}), where $p'$ is the power-law spectral index of emitting electrons. Therefore, the presence of the large-scale magnetic fields would increase the total radiation fluxes and the corresponding polarization degree. Similar to a lot of literature published (\citealt{Veledina13,Zhang14,Romero14,Zdziarski12,Zdziarski14,Zdziarski16}), which presented an emission possibility originating from accretion disk or corona region, we give up fitting the MeV tail data using the jet model. In the fittings, we return to more commonly used parameters, such as $p=2$ and $\alpha=5/3$. The results show that the model also predicts a high polarization degree at high-energy $\gamma$-ray region ($<10^{23}$ Hz).

Concerning turbulent magnetic fields, we consider that the variations of the turbulent magnetic field strength are distributed as a function of the height $z$ along the jet. In addition to the spatial correlation of magnetic field, they are assumed to be isotropic and homogeneous in the plane compressed by shocks, for the sake of simplicity. In the simulation of synthetical data cube of MHD turbulence, it is usually regarded as a Gauss type distribution with narrow dispersion (e.g., \citealt{Zhang16}). In some case, a power-law form with wide dispersion is also used (e.g., \citealt{Kelner13,Prosekin16}). Besides, in the environment mixing large- and small-scale turbulent magnetic fields, their relative strengths are still needed to further study theoretically.

Our study has assumed that the compressed slab of turbulent magnetic fields is in the same direction along the whole dissipation region of the jet. To a large extent, the `slab', which is produced by compression of shock waves and shearing at the jet boundary layer of initially chaotic magnetic fields, should be symmetrically oriented around jet axis. The simplest geometry is that the normal to the slab \textbf{\emph{s}} is parallel or perpendicular to the jet axis \textbf{\emph{j}}. The former corresponds to $\vartheta=0^\circ$, that is, the plane direction of the slab is perpendicular to \textbf{\emph{j}}, whereas the latter to $\vartheta=90^\circ$. More complicated (also realistic) configuration is that the configuration of slabs with the normal vector \textbf{\emph{s}} is at some fixed angle to \textbf{\emph{j}}, so that directions of \textbf{\emph{s}} would create a cone around \textbf{\emph{j}}, i.e., conical geometry of turbulent magnetic fields compressed. The study of these complicated configurations is beyond the scope of the paper, but it is expected that they would result in more significant polarization degree, which in local region may be changing away the jet axis and reach maximum possible value at the edges of the jet, due to more anisotropic turbulent structures. In this case, the resulting directions of linearly polarized radiation on the sky would also change along with the angle between \textbf{\emph{j}} and \textbf{\emph{s}}.

The results of the model application to Cygnus X--1 show that the angle between the line of sight and the normal to the slab of magnetic fields is $\sim75^\circ$, which implies that the angle between the direction of the jet axis and the normal direction to the slab plane is $\sim47^\circ$, using observational polarization angle at MeV tail $\sim40^\circ$ (\citealt{Rodr15}) and the position angle of the jet $\sim 23^\circ$ (\citealt{Stirling01}). Otherwise, if the position of the normal of the slab $\varphi_{\rm pos}$ is determined a priori, one can reproduce observational polarization angle. In addition to the presence of the highly polarized $\gamma$-ray emission, Cygnus--X emits also low polarization radiation in radio, IR, optical, UV and X-ray wavebands. The purpose of the current work is focused on the highly polarized $\gamma$-ray emission under the constraint of radio and IR observations. This work predicts a high polarization degree, $\sim30\%$, in radio, IR and optical bands. It should be noticed that our studies only take into account an intrinsic polarization emission from the microquasar jets. In the realistic scenario, several depolarization processes, such as internal (external) Faraday dispersion, gradients in rotation measurement across the telescope beam, a reduced polarization due to unpolarized radiation components, and the self-absorption effect at low-frequency bands, could be at work to reduce the intrinsic polarization degrees predicted in the work. Hence, although the model has predicted the degree of polarization of about $10\%$ at high-energy $\gamma$-ray region (see Figures \ref{figs:fittingtail} and \ref{figs:fittingGeVjitter}), it is unlikely to detect a polarization signal in the $>10^{23}$ Hz ranges.
From this point of view, $\varphi\approx75^\circ$ should be a lower limit required for explaining MeV tail observations.

The lowly polarized UV and X-ray observations are usually explained by interstellar dust emission component and thermal plasma corona radiation surrounding the central black hole, respectively (e.g., \citealt{Laurent11,Russell14}); this work does not explore these points. Our work argues that a high polarization degree does not necessarily require the field to be uniform in the jet of X-ray binaries, and claims that it is important for the presence of small-scale turbulent magnetic fields. In the next years, \emph{Fermi} LAT has the potential to detect high degrees of polarization from some of the bright $\gamma$-ray binaries (\citealt{Giomi16}). The polarization observation at high energy $\gamma$-ray region ($<10^{23}$ Hz), which can break degeneracies between radiative mechanism in theoretical models, is an excellent way to test our theory model. Even so, many important points, such as turbulence spectral slope, magnetic field structure, distributions of relativistic electrons, and electron acceleration mechanism, have an urgent need for in-depth study.

\acknowledgments
We would like to thank the anonymous referee for constructive suggestions and comments that significantly improved our manuscript. We are indebted to financial support from the National Natural Science Foundation of China under grants 11233006, 11363003, 11273022 and 1531108. J.F.Z. acknowledges support from the research project of Xiangtan University under grant No. KZ08089.

\end{document}